\documentclass[a4paper]{jpconf}

\usepackage{graphicx}
\usepackage{amsmath,amssymb}
\usepackage{slashed}

\newcommand{\gfcc}[2]{\frac{#1\pi G}{c^{#2}}}
\newcommand{\teq}[1]{\hbox{$#1$}}

\newcommand{\Lm}{\mathcal{L}}
\newcommand{\Em}{\mathcal{E}}
\newcommand{\Um}{\mathcal{U}}
\newcommand{\Tm}{\mathcal{T}}

\newcommand{\nline}{\nonumber \\ }

\newcommand{\detg}{\sqrt{-g}}

\newcommand{\dpar}[2]{\frac{\partial #1}{\partial #2}}

\begin{document}

\title[What quantum matter tells about quantisation of gravity]{What quantum matter tells about quantisation of gravity in a statistical mechanics context}

\author{Pierre A Mandrin}

\address{Independent researcher, Glarnerstrasse 53a, 8854 Siebnen, Switzerland}

\ead{mandrin@hispeed.ch}

\begin{abstract}
We propose a ``guide'' towards quantisation of gravity based on quantum matter in a statistical mechanics context. On one hand, a statistical mechanics model naturally arises from the thermodynamic interpretation of horizons in Rindler space. On the other hand, the path integral formulation of quantum field theory can be interpreted from the point of view of statistical mechanics. From these perspectives, gravity and matter are related to each other in the same way as a gas and its chemical potential are. This statistical mechanics interpretation of gravity and matter suggests that gravity should be quantised in a precise way which is determined by the quantisation of matter. Although, in a first step, quantisation of gravity applies for small perturbations of the metric with respect to the vacuum, the most central and general features of quantisation (quantisation prescription, quantum space dimensions) are supported from statistical mechanics and remain valid non-perturbatively.
\end{abstract}

\section{Introduction}
\label{intro}

There is, currently, a large uncertainty on how to treat gravity in a quantum mechanics context. Although it is mainly believed that gravity is required to be quantised, it is at least unclear which type of quantisation mechanism should be used, or which variables should be the preferred ones to start with. To a large part, this situation is certainly due to a lack of available experimental data which could give insight to near-Planck scale physics.

\paragraph*{}
Nevertheless, important hints should also be available from well-controlled knowledge of quantum matter. Indeed, the consequences are not obvious to investigate because we first need to understand the fundamental role of matter within the framework of gravity. By considering gravity via its statistical mechanics interpretation, matter acquires a new quality in the thermodynamic context, i.e. it acts in the same way as does the fugacity or the chemical potential in a grand canonical ensemble. The chemical potential interpretation of matter has already been proposed in \cite{Mandrin_2019}, but an important  new development step is given by identifying the grand canonical ensemble which explicitly deals with open systems both with respect to ``gravitational particles'' and an ``energy-like'' extensive quantity which comes from the Lagrangian. With the grand canonical ensemble interpretation, the quantisation of gravity can first be obtained in the perturbative regime (linearised gravity). Due to the statistical mechanical context, important features of quantisation still survive in a non-perturbative concept, namely the path integral quantisation ansatz, the preferred canonical variables for quantisation and the space dimension 3 for gravitational quantum statistical mechanics. Despite the undeniable importance of these conceptual features, their role is not to constitute the full information for presenting a complete quantum theory of gravity. Rather, the importance of the this article relies in improving the confidence for how gravity is quantised.

\paragraph*{}
The concept of the statistical mechanics interpretation of  the boundary term of gravity is introduced in Section~\ref{stat}. It is based on an argument in \cite{Mandrin_2017}. The newer interpretation in this article has some different features; the main one is the open system property of a compact space region, thus providing to the total action the quality of a grand canonical potential. To lowest perturbation order, a matter field has the effect to linearly shift the gravitational field away from the classical vacuum solution. 
 
\paragraph*{}
There is a subtile connection between matter and gravity: the radiated matter perceived by accelerated observers (Unruh effect) is nothing but the heat (matter particles) crossing the Rindler horizon \cite{Unruh, Jacobson, Padmanabhan_2014}. An equivalent process is the Hawking radiation produced in conjunction with black hole entropy \cite{Hawking, Bekenstein, Wald}. It is therefore natural to interpret the radiated matter as a thermodynamical manifestation of gravity, although this does not mean that gravity itself is specific to or can be identified with particular species of matter or their Hamiltonians. The open system concept and grand canonical ensemble view is introduced in Section~\ref{grand}.

\paragraph*{}
In Section~\ref{quantum}, we conclude from Sections~\ref{stat} and~\ref{grand} how quantisation of gravity follows from quantisation of matter. The same type of canonical quantisation procedure as for flat space-time Quantum Field Theory (QFT) applies for linearised gravity, with the requirement that one of the canonical variables has to be the vielbein, and path integral quantisation applies in a similar way as for QFT, appart from the evolution aspect. The path integral concept for gravity and matter has a one-to-one correspondence with the statistical mechanics formulation of gravity, i.e. it is not bound to a perturbative treatment, and the vielbeins therefore must keep their role of canonical functional variable for the non-perturbative quantisation procedure. The matter part in the path integral corresponds to the term containing the chemical potential $\mu$ in the exponent of the grand canonical partition function of a gas. 

\paragraph*{}
From Section~\ref{stat}, the general form of quantisation of gravity must have its most natural formulation in 3d-space since this is the dimension of the Rindler horizon hypersurfaces which the initial microcanonical ensemble treatment is originating from. This dimension is automatically recovered in the path integral treatment when the ``external time evolution'' problem is removed. The reduced dimension 3 indeed corresponds to the requirement for a unitary quantum gravity theory \cite{Dim_Red}.

\section{Summary of the statistical mechanical interpretation of gravity}
\label{stat}

We first shortly review the required parts from \cite{Mandrin_2017} and references therein and strengthen the required chain of argumentation as well as the physical picture of the statistical mechanical interpretation of space-time. As is well-known from \cite{Jacobson, Padmanabhan_2014}, space-time has a thermodynamical interpretation. Any spatial 2-surface on any 3-dimensional null-hypersurface is perceived as a Rindler horizon by a suitably located, uniformly accelerated observer, with an associated temperature \teq{T = \kappa / (2\pi)} ($\kappa$ is the accelleration of the observer or Rindler surface gravity) and an entropy \teq{S_{hor}=A/(4 L_p^2)} related to a surface area $A$, where $L_p$ is the Planck length, and the horizon of the Rindler wedge is equivalent to a black hole horizon. It should be stressed that the unboosted original space with coordinates ($t, x, y, z$) does not need to be flat, it may, in general, have a non-singular metric \teq{g_{\alpha\beta}\ne 0,\pm\infty}. By the generalised terminology ``Rindler space'', we therefore mean the boosted space defined by the coordinates ($T, X, y, z$),

\begin{equation}
\label{Rindler}
t = X \sinh T, \qquad x = X \cosh T.
\end{equation}

\noindent According to \cite{Majhi_Padmanabhan, Padmanabhan_2014}, a 3-dimensional null-hypersurface in Einstein gravity contributes a surface action $\mathcal{A}_{sur}$ (which corresponds in part to the boundary term), and $\mathcal{A}_{sur}$ is interpreted as the time integrated ``heat'' \teq{\int dt \ Q = \int dt \ TS} when evaluated on the Rindler horizon (taking care of the order of evaluation steps for taking the null limit). We have: 

\begin{equation}
\label{S_d1_P}
\mathcal{A}_{sur} = \frac{c^4}{16 \pi G} \int N^c_{ab} \ f^{ab} \ d^3\Sigma_c, \qquad f^{ab} = \sqrt{-g} \ g^{ab}, \qquad N^c_{ab} \ d^3\Sigma_c = N^{ab} \ d^3x,
\end{equation}

\noindent where, in the notation of \cite{Padmanabhan_2014}, $f^{ab}$ is the entropy density, $N^c_{ab}$ is its conjugate momentum, and $d^3\Sigma_c$ is the 3-surface element covector. On the right hand side of (\ref{S_d1_P}), \teq{N_{ab} = K_{ab} - K^c_c \gamma_{ab}} and $K_{ab}$ is the second fundamental form \cite{Gibbons_Hawking, York}.  The interpretation of (\ref{S_d1_P}) and of the elaborations of \cite{Jacobson, Padmanabhan_2014} is that every spatial surface represents a ``gas'' of gravitational building blocks (``gravitational particles''). Notice that, in order to keep the entropy and thus the sample of these ``particles'' finite while avoiding infinitely long observer orbits, it is preferable to integrate over a finite section of 3d-hypersurface. However, this must be handled with care since cutting a finite section out of an infinite hypersurface yields an open system, we consider this in more detail in the next section.

\paragraph*{}
The basic idea of our effort is that, whenever a space-time region possesses a boundary surface and we compute the physical content by varying the bulk gravitational action for this space-time region, we omit to vary the divergence term integrated over the space-time part beyond the boundary surface, and this missing information content is equivalent to the missing content that we suffer by placing a Rindler horizon exactly in that boundary surface and observing the Unruh radiation. In order to obtain a one-to-one correspondence, the observer must choose the accelleration $\kappa$ so that the hidden heat contents of both frames match. In the frame of original space, the omitted part of the action is a surface contribution of the boundary term $\Delta S_{\partial V}$, and this is equal to \teq{\mathcal{A}_{sur}=-\int dt H_{sur}} in the language of \cite{Majhi_Padmanabhan}, provided that a proper gauge of the metric is chosen. The boost accelleration $\kappa$ must be chosen in such a way that \teq{H_{sur}} in original space is equal to its value \teq{TS} computed in the Rindler frame, i.e. $T$ must have the same value in both frames since the surface areas in the \teq{yz}-subspace (i.e. the entropies) are equal. There remains an issue that 3-surfaces of boundary terms are not null. However, as shown in \cite{Mandrin_2017}, any non-null section of smooth 3d-hypersurface in the boundary term integral can be replaced to any desirable accuracy by a sequence of narrow sections of null-surfaces (strips) glued together (much like a smooth ramp would be approximated by narrow stairs), by replacing \teq{\int_{\Delta x \Delta A} \rightarrow \int_{\Delta u \Delta A} + \int_{\Delta v \Delta A}} (the strip widths $\Delta u$, $\Delta v$ being the null decomposition of $\Delta x$). Consider now a compact space-time region $V$ with piece-wise smooth boundary \teq{\partial V = \sum_A B_A}, where every component $B_A$ is non-null (time- or space-like). Then, the boundary term is the sum of the time integrated heats as perceived by all the accelerated observers related to the null-strips covering the boundary $\partial V$:

\begin{equation}
\label{S_dV_heat}
S_{\partial V} = -\int dt \sum_k T_k S_k |_{\Delta\partial V_k} ,
\end{equation}

\noindent where $\Delta\partial V_k$ are small sections covering \teq{\partial V = \bigcup_k\Delta\partial V_k} with the associated temperatures $T_k$ and entropies \teq{S_k = \ln \Omega_k}, and $\Omega_k$ is the number of distinct microstates which are compatible with the macroscopic state of $\Delta\partial V_k$.

\paragraph*{}
For theories of gravity beyond Einstein gravity, the definition of temperature and heat can be redefined in such a way that the interpretation of the boundary term as  in (\ref{S_dV_heat}) is preserved, while the integrands of the null-strip integrals will normally undergo a higher order modification (in terms of the metric), depending on the theory. In this sense, our statistical mechanical considerations are not restricted to Einstein gravity.

\section{The grand canonical ensemble interpretation}
\label{grand}

\paragraph*{}
Expression (\ref{S_dV_heat}) is not best suited to describe finite sections of 3-surfaces. Since the bounds of a section are not bounds of an autonomous physical system, there must be an uncertainty about the number of ``building blocks'' of gravity that are located inside of a section. Even more, since quanta of matter are subject to the Heisenberg uncertainty relation and thereby affect the gravitational field so as to make it blurred, the ``gravitational particles'' may hardly be localisable, and a section may contain anything between zero and all of the ``gravitational particles''.  A finite section therefore represents an open thermodynamic system with respect to the exchange of ``gravitational particles'' and also with respect to the exchange of independent extensive quantities associated with the ``gravitational particles'' -- we will find such one (much like the ``energy'' of a molecular gas), and the part of the 3-surface outside the section is a particle bath and a ``thermal'' bath. The suitable ensemble is therefore the grand canonical ensemble for which the density of states is formally given by

\begin{equation}
\label{rho_grand}
\rho_i = \frac{\exp{[(-\Em_i + \mu N_i)/(k_BT)}]}{Z}
\end{equation}

\noindent with the grand canonical partition function

\begin{equation}
\label{PF_grand}
Z = \sum_i\exp{[(-\Em_i + \mu N_i)/(k_BT)]}.
\end{equation}

\noindent In (\ref{PF_grand}), the sum is over a phase space partition with (infinitesimal) cell label $i$. Since ($\ref{rho_grand}$) does not represent a classical ``molecular gas'', the symbol $\Em$ may be a quantity different from the ``energy''. The precise interpretation of the $\Em_i$ and ``chemical potential'' $\mu$ is non-trivial and cannot be derived from the 3-surface consideration of Section~\ref{stat}, but can be identified via general thermodynamic properties of the ensemble. A key quantity is the grand canonical potential \teq{\Phi = k_BT\ln{Z}} which is minimised in the state of thermodynamic equilibrium. Taking the logarithm of (\ref{PF_grand}) yields

\begin{equation}
\label{Phi_grand}
\Phi = \Um - TS - \mu N,
\end{equation}
 
 \noindent where  \teq{\Um=\langle \Em\rangle=\sum_i\rho_i \Em_i} is the ensemble mean of $\Em$, likewise \teq{N=\langle N\rangle=\sum_i\rho_i N_i}, and \teq{S=\langle -k_B \ln\rho\rangle=\sum_i -k_B \rho_i \ln\rho_i}. In terms of dynamics, the quantity which must be extremised is the gravitational action $S_g$ plus boundary term $S_{g\partial V}$, supplemented by the action $S_m$ for matter plus boundary term $S_{m\partial V}$, this altogether must be proportional to $\Phi$ in order for (\ref{S_dV_heat}) to be satisfied under general conditions:

\begin{equation}
\label{Phi_grand_total_action}
-t\Phi = S_g + S_{g\partial V} + S_m + S_{m\partial V}.
\end{equation}
 
\noindent The term $S_{m\partial V}$ is required if matter contributes on the boundary, which is expected for compact~$V$. For instance, a real scalar field has the boundary term \teq{-\int_{\partial V}d^3x\sqrt{-g}\varphi\nabla_\perp\varphi}. As indicated by the sum of expression (\ref{S_dV_heat}), $T$ and $\mu$ may be allowed to vary between small 3d-volume elements. However, these variations must evolve slowly, i.e. the variation (wave) length scale e.g. of $T$ should be much larger then the mean free path $\lambda$ as given from interaction theory (such lengths are expected to be much smaller than experimentally achievable spatial resolutions), according to standard kinetic theory of gases. Accordingly, (\ref{PF_grand}) should be extended to

\begin{equation}
\label{PF_grand_int}
Z = \sum_i \exp{\bigg[\int_{\Sigma}(-\epsilon_i + \mu \eta_i)/(k_BT)\bigg]},
\end{equation}

\noindent where the sum over 3-volume elements has been converted to an integral, $\epsilon_i$ and $\eta_i$ are 3d-densities of $\Em_i$ and $N_i$, respectively, and $\Sigma\subset\partial V$ is the space-like boundary section on the past 3-surface). Then, (\ref{Phi_grand_total_action}) becomes:

\begin{equation}
\label{Phi_grand_total_action_int}
\int dt \int_{\Sigma} \big[ -\epsilon + Ts + \mu \eta\big] = S_g + S_{g\partial V} + S_m + S_{m\partial V},
\end{equation}
 
\noindent where $s$ is the 3d entropy density. Now, whenever we vary the metric both in $V$ and on $\partial V$, the effect is to change the values $S_g$ and $S_{g\partial V}$. Notice that $s$ and $\eta$ are independent variables (since heat and ``particle'' density vary independently from each other), i.e. we can have a variation $\delta s$ for fixed $\eta$. Minimisation of $\Phi$ only implies a change $\delta\epsilon$, i.e. we can separate, in this scenario, the gravitational part of (\ref{Phi_grand_total_action_int}) from the matter part:

\begin{eqnarray}
\label{Phi_grand_grav_action_no_mu}
\int dt \int_{\Sigma} \big[ -\epsilon + Ts \big] & = & S_g + S_{g\partial V}, \\
\label{Phi_grand_mat_action_mu}
\int dt \int_{\Sigma} \mu \eta & = & S_m + S_{m\partial V}.
\end{eqnarray}
 
\noindent As we see, the density of $H_{sur}$ reappears in $Ts$ and is also contained in $S_{g\partial V}$, therefore $\epsilon$ must contain the (densitised) gravitational bulk Lagrangian. It is also apparent that the term $S_{g\partial V}$ acts as a Legendre transformation since the variable to be varied on the boundary changes (\teq{g_{ab}\rightarrow \Gamma^c_{ab}}) and the left-hand-side term $Ts$ performs the same Legendre transformation in thermodynamic language (\teq{S\rightarrow T}), i.e. $\epsilon$ is precisely the bulk Lagrangian density. From (\ref{Phi_grand_mat_action_mu}), $\mu$ represents the (densitised) matter Lagrangian per particle (including the divergence contribution constituting the boundary term). There is also another way to see that the $\mu\eta$-term represents matter: A variation $\delta\eta$ around \teq{\eta=0} causes a change \teq{\delta(\epsilon - Ts)} upon minimisation of (\ref{Phi_grand_total_action_int}), which modifies the physical metric in $V$ such that the vacuum equations of gravity are violated. In the Einstein gravity limit, we only vary the metric, and it is required to add a matter term and vary the matter field in order to restore thermal equilibrium, hence the $\mu\eta$-term corresponds to the matter contribution to the action.
 
\paragraph*{}
In order to deepen our understanding of the role of matter in the context of gravity, it is useful to consider linearised gravity with matter. For any Lagrangian theory of gravity, the general field equations are

\begin{equation}
\label{eq:grav_eq}
\delta(\detg \Lm_g) = -\gfcc{8}{4}\delta(\detg \Lm),
\end{equation}

 \noindent where $\Lm_g$ and $\Lm$ are the gravitational and matter Lagrangian, respectively. We linearise (\ref{eq:grav_eq}) and neglect sources for torsion (or for any other extra-variables) for consistency, thus obtaining the linearised Einstein equations with stress tensor $T_{\alpha\beta}$:

\begin{equation}
\label{eq:Cl_Rab_lin_Einstein}
g_{\alpha\beta,\mu,}{}^\mu +  g_{\mu\nu,\alpha,\beta}\eta^{\mu\nu} - g^\mu_{\ \alpha,\mu,\beta} - g^\mu_{\ \beta,\mu,\alpha} = -\gfcc{8}{4} \bigg[T_{\alpha\beta}-\frac{T}{2}\eta_{\alpha\beta}\bigg],
\end{equation}

\noindent where the cosmological constant has been omitted. In linearised theory, it is sufficient to neglect couplings between interacting matter fields and to take into account terms up to second order only in the matter field for $T_{\alpha\beta}$. In the scope of the standard model,

\begin{equation}
\label{eq:stress}
T_{\alpha\beta} = 2 \dpar{\Lm}{g^{\alpha\beta}} - \Lm g_{\alpha\beta}
\end{equation}

\noindent and $\Lm = K - V$ with the kinetic term $K$ and the potential term $V$ (\teq{K, V} each admit one contribution per species). We evaluate $\partial\Lm/\partial g^{\alpha\beta}$, then eliminate all $V$-contributions to $T_{\alpha\beta}$ via the equations of motion of matter, e.g.

\begin{equation}
\label{eq:L_fromK}
\Lm= \alpha \nabla_\mu (\dpar{K}{\nabla_\mu \varphi} \varphi) - \beta \dpar{K}{\varphi} \varphi - \gamma (\nabla_\mu \dpar{K}{\nabla_\mu \varphi}) \varphi
\end{equation}

\noindent is the Lagrangian for every single species, where e.g. \teq{\alpha=1/2, \beta=\gamma=0} for a neutral scalar field (\teq{\varphi = \varphi^\dagger}), and supplementing an index (\teq{\varphi\rightarrow \varphi^\mu}) in the case of a vector field and like-wise  in any later expressions. For an (approximately) monochromatic field (single-k-mode idealisation for linearised theory) of a single species with Fourier transform $\tilde{\varphi}(k^\mu)$, a straight-forward computation yields that it is proportional to the tetrad perturbation \teq{h^I_{\ \ \alpha}=e^I_{\ \ \alpha}-\delta^I_\alpha} in the neighbourhood of a space-time point and the non-oscillatory background frame is chosen to be locally Minkowski:

\begin{equation}
\label{eq:subst_uphi_g}
\tilde{h}^I_{\alpha} = \vartheta_{\alpha}^I \tilde{\varphi}, \qquad g_{\alpha\beta} = h^I_{\alpha} \ \eta_{IJ} \ h^J_{\beta} + g_{0\alpha\beta}.
\end{equation}

\noindent E.g. for a neutral scalar field, we have 

\begin{eqnarray}
h^I_\alpha & = & -2 \sqrt{\frac{\pi L_p^2}{\hbar}} \ \delta^I_\alpha  \ \varphi_0 \sin(k_\mu x^\mu), \nline
\label{eq:h_sq_main}
g_{0\alpha\beta} & = & \eta_{\alpha\beta} - 2 \frac{\pi L_p^2}{\hbar} \ \eta_{\alpha\beta}   \ \varphi_0^2
\end{eqnarray}

\noindent with plane wave field amplitude $\varphi_0$. (\ref{eq:subst_uphi_g}) may be extended to admit severel species $l$ by replacing \teq{g_{\alpha\beta}\rightarrow \sum_l g^{(l)}_{\alpha\beta}} and writing the proportionality for each contribution $h^{(l)I}_{\ \ \alpha}$. Despite the special limit, (\ref{eq:subst_uphi_g}) reveals that the presence of matter acts as a shift constraint for gravity, i.e. the matter field must be compensated by a shift of the tetrad. This translates to the thermodynamics view according to which a change $\delta N$ of particles causes a shift $\delta \Em$ in order to keep the ``gas'' in equilibrium (minimisation of $\Phi$). At least in the monochromatic linearised case, the particle number $N$ corresponds to the number of building blocks of gravity to be generated in order for the matter field to emerge. Due to the fundamental nature of the thermodynamics concept, the basic interpretation of $N$ should persist even when nonlinear corrections to (\ref{eq:subst_uphi_g}) gradually become relevant.

\section{From quantum matter to the quantisation of gravity}
\label{quantum}

From the above considerations, the properties of quantum matter in the flat space-time approximation allow to infer certain properties of gravity at microscopic level. In the flat space-time approximation, the partition function (\ref{PF_grand_int}) simplifies:

\begin{equation}
\label{PF_grand_flat}
Z_{flat} = \sum_i \exp{\bigg[\int_{\Sigma}\mu \eta_i/(k_BT)\bigg]},
\end{equation}

\noindent $Z_{flat}$ corresponds to the path integral for matter for a space-time ``region'' with spacelike boundary contributions in the form of time slices at \teq{t=0} and \teq{t=\Tm}; we consider one species with label $l$:

\begin{equation}
\label{eq:path_m}
Z_{flat}^{(l)}\big|_{0\rightarrow \Tm} = \langle\varphi^{(l)}(t=\Tm,x^i)|e^{-iH\Tm}|\varphi^{(l)}(t=0,x^i)\rangle = \int \mathcal{D} \varphi^{(l)} \ e^{i\int_0^\Tm d^4x\Lm[\varphi^{(l)}]/\hbar}.
\end{equation}

\noindent (\ref{PF_grand_int}) together with (\ref{PF_grand_flat}) and (\ref{eq:path_m}) imply that the gravitational extension must be of the form

\begin{equation}
\label{eq:path_g_ansatz}
Z\big|_{0\rightarrow \Tm} = \int \ f(X) \mathcal{D} X \prod_l \mathcal{D} \varphi^{(l)} \ e^{i[S_{grav\Delta V}+S_{grav\Delta\partial V}]_{0\rightarrow \Tm}/L_p^2 + i\sum_l[S^{(l)}_{m V}+S^{(l)}_{m \partial V}]_{0\rightarrow \Tm}/\hbar},
\end{equation}

\noindent where $X$ is the (not yet determined) gravitational functional variable which is related (but does not need to be equal) to the thermodynamic quantity $\epsilon$, the function \teq{f(X)} allows for an adaptation factor which has the value $1$ in  flat space-time approximation; $f$ may be necessary at least to ensure invariance of the density of states under gauge transformations. The fixed time limits \teq{0,\Tm} are still present as a parameter to use for time evolution. However, the phase space over which the ensemble sum is taken departs from the original idea in (\ref{eq:path_m}), namely that the integral should have a common time interval, possibly an observable temporal distance \teq{\langle\sqrt{\Delta x^\mu g_{\mu\nu}\Delta x^\nu}\rangle} and not a meaningless coordinate time. Moreover, it is not clear, in the context of a covariant theory of gravity, why one should evolve paths along time rather than along space. One possible modification would be to represent all the space-time orientations in the sum over path evolutions. However, this would require to reduce the degrees of freedom afterwards in order to recover (\ref{eq:path_m}), and the problem of ensemble averaged evolution parameter would persist. These concerns may directly be avoided if we remove the time limit and abandon the time dimension from the beginning. We shall come back to this idea later. 

\paragraph*{}
From (\ref{eq:path_g_ansatz}) and from the point of view of the present statistical mechanical interpretation, (\ref{eq:path_m}) does not represent a canonical partition function since it is reminiscent of the $\mu N$-part of (\ref{eq:path_g_ansatz}). In the form (\ref{eq:path_g_ansatz}), the gravitational path integral already reveals a basic consequence, namely that the gravitational field must be quantised in order for the matter field to be quantised, since matter plays the role of fugacity within gravity. Starting with classical statistical mechanics on the gravity side would make quantum matter ($\mu N$-part) incompatible. More precisely, when the particle number is promoted to a quantum operator $\hat{N}$, with generally mixed states in particle number representation, gravity itself must have mixed states in terms of its number of building blocks, via the relation (\ref{eq:subst_uphi_g}), i.e. we must introduce operators $\hat{e}^I_{\alpha}$ and therefore $\hat{X}$. The properties of gravitational quantum mechanics is even forced to be compatible with quantum mechanics of matter since, for linearised theory, the transition functions with time slices \teq{t=0,\Tm} for the boundary exhibit the same formal behaviour for gravity and matter, as suggested by the following symmetric notation and admitting a single species:

\begin{eqnarray}
& \langle X(0,x^i),\varphi(0,x^i)|e^{-iH\Tm}|X(\Tm,x^i),\varphi(\Tm,x^i)\rangle & \nline
\label{eq:path_g_lin}
& = \int \ \mathcal{D} X \mathcal{D} \varphi \ e^{i\int_0^\Tm d^4x\sqrt{-g}\Lm_g/L_p^2 + i\int_0^\Tm d^4x\sqrt{-g}\Lm/\hbar}. &
\end{eqnarray}

\noindent In linearised theory, the path integral quantisation (\ref{eq:path_g_lin}) is equivalent to canonical quantisation:

\begin{equation}
\label{eq:X_comm}
[\hat{\Pi}(x^a,t),\hat{X}(y^a,t)]  \sim  L_p^2 \delta(x^a-y^a), \qquad [\hat{\pi}(x^a,t),\hat{\varphi}(y^a,t)]  =   i \hbar \delta(x^a-y^a),
\end{equation}

\noindent where $\hat{\Pi}$ and $\hat{\pi}$ are the momenta canonically conjugate to $\hat{X}$ and $\hat{\varphi}$, respectively. Using (\ref{eq:subst_uphi_g}) and the symmetry \teq{X\leftrightarrow \varphi} of (\ref{eq:path_g_lin}), we see that any variation $\delta\tilde{\varphi}$ modifies in the exact same way the weights in the sum of paths as does the variation of its counter-part \teq{\delta\tilde{e}^I_{\alpha} = \vartheta_{\alpha}^I \delta\tilde{\varphi}}, i.e. if we set \teq{X=e^I_{\alpha}}, both variables will generate the exact same patterns of transition matrix elements (\ref{eq:path_g_lin}), i.e. both variables have the same states ($|\varphi\rangle$ and $|e^I_\alpha\rangle$) and commute. This is also equivalent to writing the operator version of (\ref{eq:subst_uphi_g}):

\begin{equation}
\label{eq:subst_uphi_g_q}
\hat{\tilde{h}}^I_{\alpha} = \vartheta_{\alpha}^I \hat{\tilde{\varphi}}.
\end{equation}

\noindent As an example, a real scalar with plane wave amplitude \teq{\varphi_0=\sqrt{\hbar c^2 / (2V_3\omega_k')}} will yield

\begin{equation}
\label{eq:qh_sq_main}
\hat{h}{}^I_\alpha = \sqrt{\frac{\pi L_p^2}{\hbar}} \ \delta^I_\alpha  \ \varphi_0 \left(-ie^{-ik_\mu x^\mu}\hat{a}_{k^i}+ie^{ik_\mu x^\mu}\hat{a}^\dagger_{k^i}\right).
\end{equation}

\noindent The first quantisation prescription in (\ref{eq:X_comm}) is thus satisfied if we identify \teq{\hat{X}\rightarrow \hat{h}^I_{\alpha}} or \teq{\hat{X}\rightarrow \hat{e}^I_{\alpha}}. In linearised theory, the following simple commutation relations apply:

\begin{eqnarray}
\label{eq:qe_sq_prescr}
\left[\hat{e}{}^I_\alpha(x^i,t),\hat{e}{}^J_{\beta,0}(x'^i,t)\right] & \sim & L_p^2 \delta^I_\alpha\delta^J_\beta \delta(x^i-x'^i),\\
\left[\hat{e}{}^I_\alpha(x^i,t),\hat{e}{}^J_\beta(x'^i,t)\right] & = & \left[\hat{e}{}^I_{\alpha,0}(x^i,t),\hat{e}{}^J_{\beta,0}(x'^i,t)\right] = 0,
\end{eqnarray}

\noindent where the $\delta$-function is restricted to 3 dimensions. Since we assumed vanishing torsion, (\ref{eq:qe_sq_prescr}) can be obtained directly from the right hand side of (\ref{eq:X_comm}). For comparison, the technique equivalent to matter quantisation is to use the first derivative Einstein Lagrangian \teq{\Lm_E\sim g^{\alpha\beta}(\Gamma^\gamma_{\alpha\delta}\Gamma^\delta_{\gamma\beta}-\Gamma^\gamma_{\alpha\beta}\Gamma^\delta_{\delta\gamma})} \cite{York} and then to compute the momentum \teq{\partial(\detg\Lm_E)/\partial e^I_{\alpha,0}} canonically conjugate to $e^I_{\alpha}$, a lengthy but straight-forward calculation yields an expression of the form \teq{\Pi_I^\alpha = F_{IJ}^{\alpha\beta\gamma}(e^K_\mu) e^J_{\beta,\gamma}}, where $F_{IJ}^{\alpha\beta,\gamma}(e^K_\mu)$ is a tensorial function of the tetrad and its inverse. One could wonder why we should not simply use the right hand side of (\ref{eq:X_comm}) which seems to work for linearised quantisation. However, as soon as some torsion is present, the right hand side of (\ref{eq:X_comm}) provides species-dependent and thus inconsistent commutators. Moreover, the symmetry of (\ref{eq:path_g_lin}) clearly requires that we must use the calculation method by canonical conjugation using $\Lm_E$. If we extend the theory to admit some torsion (e.g. caused by the spin of a fermion species \cite{Hehl}), the connection with contorsion in $\Lm_E$ will naturally lead to a torsion contribution to $\Pi_I^\alpha$.
For the full gravity variable $X$, one may want to consider that $e^I_{\alpha}$ might have to be multiplied by a function $\epsilon(e)$ which would be unity in linearised theory, with \teq{e=\sqrt{-g}}, yielding the operator \teq{\hat{X} = (\epsilon(e) e^I_{\alpha})^{\hat{ }}}. However, since \teq{\epsilon(e) e^I_{\alpha} \sim \varphi^*=\epsilon(\varphi)\varphi} (formally) at lowest perturbation order, the resulting commutation relation \teq{[\hat{\pi}^*, \hat{\varphi}^*]} for matter would be incompatible with the right hand side of (\ref{eq:X_comm}) or not even exist; this prevents us from allowing an extra factor $\epsilon(e)$ for $X$. The only way to escape from this restriction would be to redefine the matter field as \teq{\varphi^*=\epsilon(e)\varphi}, but this would require a complicated and not intuitive matter Lagrangian.

\paragraph*{}
We turn back to the problem of time limits in (\ref{eq:path_g_ansatz}) and procede with the removal of the fixed time limits from (\ref{eq:path_g_ansatz}) and dimensional reduction. We first have to get rid of the time dynamics. This is done by choosing $\Tm$ small, i.e. \teq{||h^I_{\alpha,0}||c\Tm/||h^I_\alpha||\ll 1}.The integration over time can then be replaced by a constant factor $\Tm$, and only the timelike parts of the boundary terms contribute. Now that the system has lost the time dynamics, we can reformulate it explicitly in 3 dimensions using the ADM-decomposition and remove the components $g_{0\alpha}$ from our system, thus reducing at once the gauge degrees of freedom. E.g. in the case of Einstein gravity, we apply the Gauss-Codazzi equations:

\begin{equation}
\label{eq:path_G_3}
Z\big|_{\tau_0}  = \int f(e) \mathcal{D} e^I_{\ \ a} \prod_l \mathcal{D} \varphi^{(l)} \ 
e^{i\tau_0 (\int_{\Sigma_f} d^3x\sqrt{e} [\mathcal{R} + K_{ab}K^{ab} - K^2]/L_p^2 + \sum_l \int_{\Sigma_f} d^3x\sqrt{e}\Lm^{(l)}/\hbar + \ldots)},
\end{equation}

\noindent where \teq{\tau_0=\sqrt{g_{00}}\Delta \Tm} is a fixed small timelike distance, $\mathcal{R}$ is the 3d-Riemann curvature, $K_{ab}$ is the second fundamental form, $e^I_{a}$ are triads (with 3d-indices $I$,~$a$), \teq{e=\sqrt{\det(g_{ab})}}, and the boundary terms are indicated by periods. We shall bear in mind that the sum over geometries in (\ref{eq:path_G_3}) is subject to the constraints of the initial value problem. By removing the time limits, the gravitational path integral quantisation has naturally led us to a 3-dimensional theory. This, however, is what we have expected from \cite{Dim_Red} and from the fact that we have started with 3-dimensional statistical mechanics of the Rindler horizon.


\section{\label{sec:conclusion}Conclusions}

This article has shown that knowledge from horizon thermodynamics and reliable knowledge from quantum matter can be useful to enforce our confidence in the choice of the road-map towards quantisation of gravity. Although quite some open questions are not answered in a definitive way and although this procedure is not intended to yield a full theory with all its details, the findings nevertheless suggest the type of required quantisation procedure and the canonical variables required for the quantisation prescription, and new ways of interpretation of the physical situation are opened. As a key feature, the statistical mechanical interpretation of gravity within compact regions yields a grand canonical ensemble which is naturally related to the path integral concept originating from quantum field theory. In the gravitational context, matter is interpreted as the fugacity. This results in a natural gravitational extension of the path integral quantisation and provides a pair of (main) canonical variables, namely the triad and its conjugate momentum. Appart from the flat space limit, quantised gravity is preferably operating in 3d-coordinates so as to avoid the appearance of an unpleasant ``external'' time evolution parameter.


\end{document}